\documentclass[twocolumn,showpacs,preprintnumbers,amsmath,amssymb,aip,jcp]{revtex4}
\usepackage{graphicx}
\usepackage{dcolumn}
\usepackage{color}
\usepackage{bm}
\usepackage{float}
\usepackage{gensymb}
\begin{document}

\title{Enhanced magnetism of Cu$_n$ clusters capped with N and endohedrally  doped with Cr}
\affiliation{Department of Condensed Matter Physics and Material Sciences, S.N. Bose National Centre for Basic Sciences, JD Block, Sector-III, Salt Lake
City, Kolkata 700 098, India}
\author{Soumendu Datta}
\author{Radhashyam Banerjee}
\author{Abhijit Mookerjee}
 \altaffiliation{Visiting Faculty, Department of Physics, Lady Brabourne College, Kolkata}
\altaffiliation[also ]{Visiting Distinguished Professor, Department of Physics, Presidency University, Kolkata}

\date{\today}

\begin{abstract}
The focus of our work is on the production of 
 highly magnetic materials out of Cu clusters. We have studied the relative effects of N-capping as well as N mono-doping on the structural stability and electronic properties of the small Cu clusters using first principles density functional theory based electronic structure calculations. 

We find that the  N-capped clusters are more promising in producing  giant magnetic moments, such as 14 $\mu_B$ for the Cu$_6$N$_6$ cluster and 29 $\mu_B$ for the icosahedral Cu$_{13}$N$_{12}$ clusters. This is accompanied by   a substantial enhancement in their stability. We suggest that these giant magnetic moments of the capped Cu$_n$ clusters have relevance to the observed room temperature ferromagnetism of Cu doped GaN. For cage-like hollow Cu-clusters, an endohedral Cr-doping together with the N-capping appears as the  most promising means to produce stable giant magnetic moments in the copper clusters.
\end{abstract}

\maketitle

\section{\label{sec:intro}Introduction}

The low coordinated surface atoms of transition metal nano-clusters play a significant role in deciding their magnetic properties.
Surface adsorption of gaseous molecules  directly affects the electronic properties of the surface atoms in clusters and influences
their  overall magnetic properties. Chemisorption  of gaseous molecules on the surface of transition metal nano-clusters and its effects,
 particularly on  magnetic properties, has attracted  attention in recent experimental as well as theoretical works. Several studies
 have revealed  spectacular variations of cluster magnetism depending on the types of chemical species and the nature of interactions
 between the surface atoms and the adsorbed species. For example, the molecular beam deflection experiment by Knickelbein's group revealed
 enhanced magnetic moment for hydrogenated  Fe$_n$ clusters as compared to pristine ones, for sizes ranging $n$ = 10 $-$ 25.\cite{knickel} 
Interestingly, this result is in contrast to those for larger iron nano-particles, as well as for thin films, in which quenching of
 magnetic moment has been found upon hydrogen absorption.  Similar trend of quenched magnetic moments upon chemisorption has also been found
 in the case of Benzene-capped and CO molecule coated Co clusters,\cite{knickel2}
 whereas  mere chemisorption of Oxygen does not affect the magnetization of Co clusters at all.\cite{oxy} The adsorption induced 
magnetism for Ni-clusters exhibits rather interesting variations. As for example, the full coverage of Ni$_n$ clusters by N$_2$ molecules
 leads to the  quenching of magnetic moment, but it survives at partial coverage.\cite{ni1} Similarly, the quenching of magnetic moment results
 with a partial adsorption of NH$_3$ molecules and a complete quenching occurs when the number of NH$_3$ molecules adsorbed equals to
 the number of surface Ni atoms. Surprisingly, further adsorption leads to reappearance of magnetic moment as a consequence of the variation of Ni-N distance with coverage of the absorbent.\cite{nh3} Again, while 
the adsorptions of CO and H$_2$ on Ni clusters decrease their magnetic moments,\cite{knickel3,rsb} the adsorption of atomic O on the 
small Ni$_n$ clusters induces either enhanced or reduced magnetic moments depending on cluster size, an effect attributed to adsorption induced reconstruction of cluster structures.\cite{salahub} Besides the 3$d$ late transition metal clusters, some limited studies of chemisorption on the clusters of 3$d$ early transition metal elements, have also been reported. Like, adsorption of a single CO molecule on Sc$_n$ clusters results quenching in magnetic moment depending on the size of clusters.\cite{co@sc} In the case of V$_n$ clusters, adsorption of N$_2$ and NO, O$_2$, C$_2$H$_4$ leads to oscillation or quenching/enhancement in magnetic moments.\cite{gas@v}

Cu atom has an electronic configuration of 3$d^{10}$4$s^1$ among the 3$d$ transition metal series. It is therefore, 
characterized by a closed shell of $d$ electrons and one $s$ electron in the outermost atomic shell, similar to  simple 
alkali metal atoms. Cu clusters are, therefore, expected to share similar properties with alkali metal clusters and these 
properties are likely to be described by the shell model\cite{shell} in a first attempt. On the other hand, the energy level separation 
between the valence $s$ and $d$ levels in a copper atom is small, giving rise to significant $s$-$d$ hybridization. 
In contrast to alkali metal clusters, this $s$-$d$ hybridization also plays important role for deciding many properties of Cu.
 Moreover, its role as a magnetic impurity in a semiconductor host, has attracted special attention for designing diluted magnetic 
semiconductors and spintronic systems because of its filled 3$d$ shell and therefore of intrinsic nonmagnetic character.\cite{dms1@zno,dms2@zno} Recently, the possibility of ferromagnetism of Cu-doped GaN with a Curie temperature above the room temperature, has been reported.\cite{gan} However, the origin of this ferromagnetism is still being debated. 

In this work, we have studied the effect of N-capping as well as N mono-doping on the structural and magnetic properties of small 
Cu$_n$ clusters using first principles density functional theory (DFT) based electronic structure calculations. We find that the 
effects of N-capping are remarkable with respect to magnetic applications of the capped systems as the N-capping induces giant magnetic moments to the capped Cu$_n$ clusters. Our study reveals that this enhanced magnetic moment results from the fact that the N-capping first induces spin polarization to the Cu-atoms through hybridization of the $s$-$d$ hybridized orbitals of Cu atoms with the $p$ orbitals of the capping N atoms and couples all the atom-centered magnetic moments ferromagnetically. We also note that such mechanism of induced ferromagnetism of the capped Cu clusters plays a major role in the observed ferromagnetism of the Cu-doped GaN semiconductors.

 In addition, we propose an endohedral doping mechanism to enhance the ferromagnetism of the N-capped systems and thereby using them as building blocks for constructing highly magnetic materials. A study of the effect of capping on the structural rearrangement of the parent cluster would be an important issue if we wish to examine
 the origin of the modified properties of the capped system with respect to its parent cluster.\cite{morph_chg}  Surprisingly, it may 
sometimes lead to the formation of  cage-like hollow structures.\cite{hollow1,hollow2}  With respect to such cage structures, an 
endohedral doping process {\it i.e.} introducing dopants {\sl within} the hollow cage, is considered to be  another promising means of tailoring 
 its intrinsic magnetic character. This can be seen in various cluster-fullerene systems, as well as for doped icosahedral-like structures 
 of many metal clusters.\cite{gd2c79n,m13} It actually performs a synergistic role in most cases,  where it helps to further stabilize the 
 cage-like structure  along with engineering its magnetic properties. For example, while there is no appreciable spin density on an
 empty cage of C$_{80}$ fullerene, the cluster-fullerene Gd$_3$N@C$_{80}$ forms a stable magnetic unit with high magnetic moment which is
largely contributed by the localized $f$ electrons of the Gd atoms.\cite{gd2c79n}  Enhanced magnetic moments of a cage-like 
icosahedral M$_{13}$ clusters ( M = metal atom ) upon appropriate substitution at the central site, has been examined extensively.\cite{m13} 
Since the N capped Cu$_n$ clusters are interesting for magnetic applications, it is essential to explore the possibilities of engineering 
its magnetic properties further. We have considered a case of cage-like icosahedral  nitrogenated Cu$_{13}$ cluster and scrutinized the
 possibility of using endohedral doping by a Cr atom  for constructing materials with even higher magnetic moments. 

The interesting point of this study is that while the bare Cu$_{12}$Cr cluster  maintains an icosahedral structure and has overall zero net 
magnetic moment, both the capped systems - Cu$_{13}$N$_{12}$ and Cu$_{12}$CrN$_{12}$ possess giant magnetic moments. The maximum value is obtained 
in case of the Cu$_{12}$CrN$_{12}$ cluster. In the following, we have first given a brief outline of the background of applied methodology 
in the Section \ref{methodology}, then a detailed analysis of the results on stability and magnetism in terms of bonding characteristics 
and electronic properties in the Section \ref{results}. The paper ends with a conclusion in the Section \ref{conclu}.

\section{\label{methodology} Computational Details.} The calculations reported in this study, were 
based on DFT {\sl within} the framework of pseudo-potential plane wave method, as implemented in the Vienna ab initio Simulation Package (VASP).\cite{kresse2} We used the Projected Augmented Wave (PAW) pseudo-potential \cite{blochl,kresse} coupled with the generalized gradient approximation (GGA) to the exchange correlation energy functional as formulated by Perdew, Burke and Ernzerhof (PBE).\cite{perdew} The 3$d$ as well as 4$s$ electrons for Cu/Cr atoms and 2$p$ as well as 2$s$ electrons for N atoms were treated as valence electrons and the wave functions were expanded in the plane wave basis set with the kinetic energy cut-off of 280 eV. The convergence of the energies with respect to the cut-off value were checked. Reciprocal space integrations were carried out at the $\Gamma$ point. For the cluster calculations, a simple cubic super-cell was used with periodic boundary conditions, where two neighboring clusters were kept separated by around 12 {\AA} vacuum space, which essentially makes the interaction between cluster images negligible. Symmetry unrestricted geometry optimizations were performed using the conjugate gradient and the quasi-Newtonian methods until all the force components were less than a threshold value of 0.001 eV/{\AA}. To determine the magnetic moment of the minimum-energy structure in spin polarized scalar relativistic calculations, the geometry optimization was performed for all the possible spin  multiplicities for each structure under the approximation of collinear atomic spin arrangements, as followed in our recent works.\cite{sdatta} We  also considered different spin arrangements among the atoms for a particular spin multiplicity. After all, special care has been taken to suppress the N-N interactions among the capping N atoms, in a similar manner of obtaining steric conformations for surface ligands in case of  passivated quantum dots.\cite{morph_chg,pass}
%############## Figure 1 ##############
\begin{figure*}
\rotatebox{0}{\includegraphics[height=7.9cm,keepaspectratio]{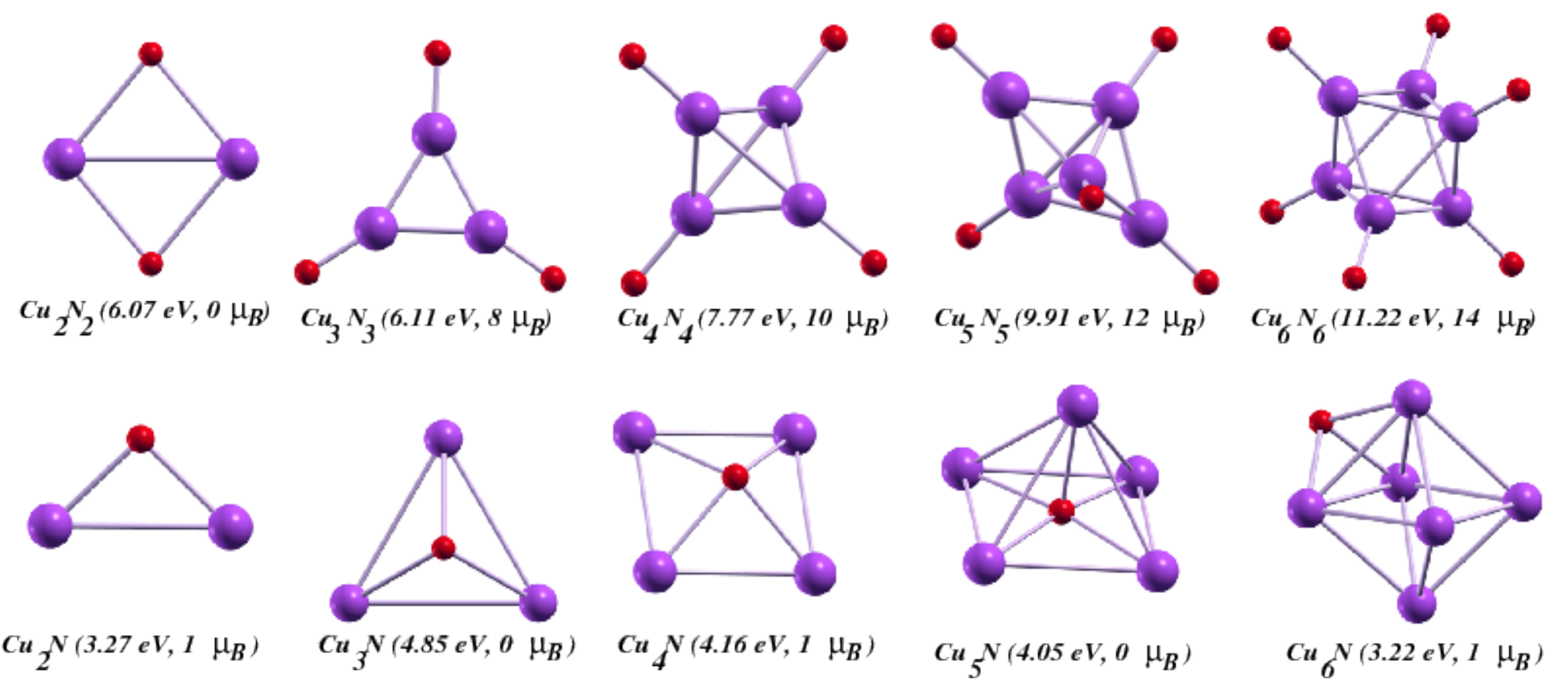}}
\caption{(Color online) Geometries of the capped Cu$_n$N$_n$  and mono-doped Cu$_n$N  clusters in their ground states for n = 2-6. The violet colored larger balls and red colored smaller balls represent Cu and N atoms respectively. The numbers in the parenthesis, represent the values of energy gain ($\Delta_c^0$ for the capped systems and $\Delta_m^0$ for the mono-doped systems) and the total magnetic moment respectively for the respective ground state.}
\label{mes}
\end{figure*}
%######################################

\section{\label{results}Results and Discussions}
We have first analyzed the optimized structures, stability and magnetic properties for the N-capped as well as N-monodoped Cu$_n$ clusters in the size range $n$ = 2$-$6 in Section \ref{capped}. We have then considered a particular case of a Cu$_{13}$ cluster and examined both the effects of N-capping as well as endohedral Cr-doping for further tuning the magnetic properties of the Cu$_{13}$ cluster. The results related to Cu$_{13}$ clusters have been discussed in the Section \ref{endo}.
\subsection{\label{capped}Capped and mono-doped Cu$_n$ clusters ($n =$ 2$-$6)}

Any theoretical study on the precise determination of the properties of an atomic cluster, first demands the determination of its minimum energy structure (MES) in complex potential energy
 surface. We have, therefore, first determined the MES for the pure Cu$_n$ clusters with $n =$ 2$-$6
 by relaxing the atomic positions and by minimizing both the total energy and spin. Several most probable initial
  configurations were tried to ensure that the optimized structure does not correspond to a local minimum. Our results indicate planar structures for each of the pure clusters in accordance with the previous reports.\cite{planar} For example, the calculated MES of the Cu$_3$ cluster, has a triangular shape and this triangular unit constitutes the building block for the MES of the subsequent higher sized clusters. Likewise, the Cu$_4$ cluster adopts a rectangular shape, consisting of two triangles, the Cu$_5$ cluster adopts a trapezoidal shape consisting of three triangles. Finally, the pure Cu$_6$ cluster has an overall triangular shape, where two interpenetrating triangular Cu$_3$ clusters are clubbed together and therefore, the composite has four triangular building blocks. Though bulk copper is nonmagnetic, small copper clusters are spin polarized and possess some week magnetic moments due to the presence of the low-coordinated surface atoms. We found that each of the even-numbered pure Cu$_n$ clusters has zero magnetic moment as all electrons are paired. On the other hand, the odd-numbered pure Cu$_n$ clusters possess a total magnetic moment of 1 $\mu_B$ each due to presence of a single unpaired electron. Note that such odd-even alterations in the magnetic moments has also been reported previously for the small pure Au$_n$ and Cu$_n$ clusters.\cite{odd-even}

Unlike the pure clusters, the capped as well as mono-doped clusters show a structural transition
 from a 2D planar to 3D geometries with increasing size for the optimized systems. Fig.\ref{mes}
  shows the MESs for the N-caped as well as N-mono doped Cu$_n$ clusters for $n =$ 2$-$6. For the 
  capped systems, the Cu$_n$ core adopts a planar geometry for $n =$2$-$3 and it is of 3D geometry for 
  $n=$4$-$6. As seen from the Fig.\ref{mes} that the ground state structures for each of the 
  Cu$_4$N$_4$, Cu$_5$N$_5$ and Cu$_6$N$_6$ clusters, contain a tetrahedral Cu$_4$, triangular bi-
  pyramidal Cu$_5$ and tetragonal bi-pyramidal Cu$_6$ cores respectively. The first isomer of the Cu$_4$N$_4$ cluster contains a rectangular Cu$_4$ core which lies 0.2 eV higher in energy from the ground state structure. Likewise, the first isomers of the Cu$_5$N$_5$ and Cu$_6$N$_6$ clusters contain a square pyramidal Cu$_5$ core and a capped triangular bi-pyramidal Cu$_6$ core respectively. Our calculated energy differences between the ground state and the first isomer of the Cu$_5$N$_5$ and Cu$_6$N$_6$ clusters, are 0.25 eV and 0.15 eV respectively. Interestingly, the isomers having a planar Cu$_n$ core, are 0.8 eV and 1.02 eV above the corresponding ground states for the Cu$_5$N$_5$ and Cu$_6$N$_6$ clusters, respectively. Table I provides more information about the structural parameters of the ground state and first isomer of the capped clusters. Note that the preferred spin multiplicities of the first isomers are the same as that of the ground state structure in case of the Cu$_4$N$_4$, Cu$_5$N$_5$ and Cu$_6$N$_6$ clusters. For the case of N-mono-doped systems {\em i.e} the Cu$_n$N clusters, the Cu$_n$ core adopts planar shape up to $n=$ 4 and it is of 3D square pyramidal and tetragonal bi-pyramidal shapes for the Cu$_5$N and Cu$_6$N clusters, respectively. For the later two mono-doped systems, the isomers having a planar Cu$_n$ core, are 0.4 eV and 0.89 eV higher in energy from the respective ground states. A detailed analysis of the ground state structures of the Cu$_n$N$_n$ clusters, reveals that the average number of nearest neighbor Cu atoms for a given Cu atom, is varying as 1, 2, 3, 3.6 and 4 for $n =$ 2, 3, 4, 5 and 6 respectively. The average value of Cu-Cu-N angles also show interesting variations along the increasing size of the capped systems, such as  $\langle Cu-Cu-N\rangle$ = 150$^\circ$ , 145$^\circ$, 140$^\circ$ and 135$^\circ$ for the optimal Cu$_3$N$_3$, Cu$_4$N$_4$, Cu$_5$N$_5$ and Cu$_6$N$_6$ clusters respectively. This uniform variation of  $\langle Cu-Cu-N\rangle$ angle establishes it as a guiding rule for the ground state geometries. On the other hand for the optimized  structure of the N-monodoped clusters, the average value of Cu-N-Cu angles varies around  90$^\circ$, such as  $\langle Cu-N-Cu\rangle$ = 91.5$^\circ$, 89.2$^\circ$, 81.7$^\circ$, 92.1$^\circ$ and  84.9$^\circ$ for the optimized Cu$_2$N, Cu$_3$N, Cu$_4$N, Cu$_5$N and Cu$_6$N clusters, respectively.
%############## Figure 2 ##############
\begin{figure*}
\rotatebox{0}{\includegraphics[height=7.9cm,keepaspectratio]{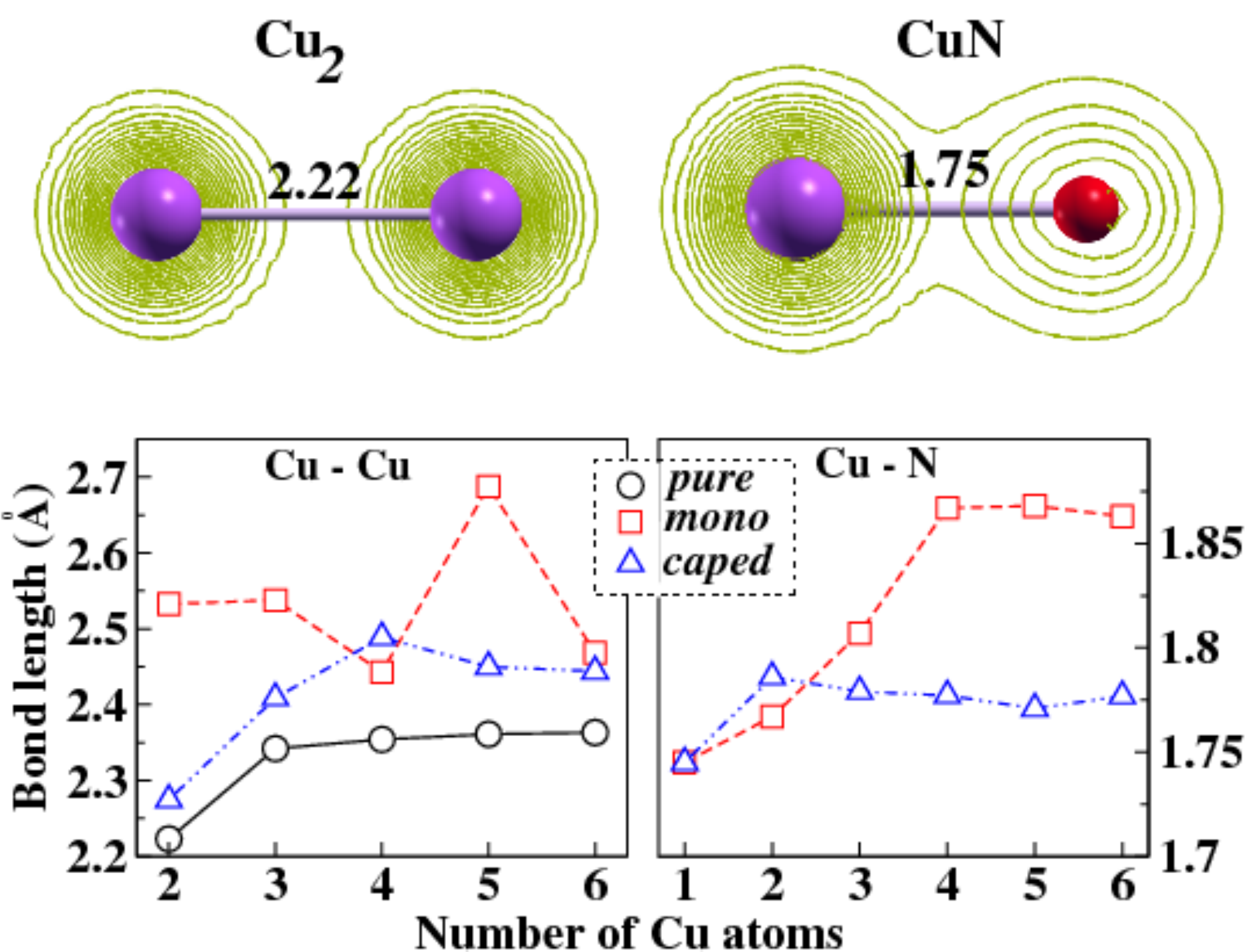}}
\caption{(Color online)Plot of charge density contour (upper panels) of Cu$_2$ and CuN dimers and variation of Cu-Cu and Cu-N bond lengths (lower panels) in the ground state structures of pure, mono-doped and caped systems. Bond lengths of dimers in the upper panel, are in Angstrom unit. }
\label{bl}
\end{figure*}
%######################################

We analyzed the energetics of the Cu$_n$N   and Cu$_n$N$_n$ clusters in terms of the total energies 
of the pure Cu$_n$ and capped/doped systems corresponding to their ground state structures. For the 
capped systems {\em i.e} Cu$_n$N$_n$ clusters, we have calculated the energy gain, $\Delta^0_c$ in 
adding $n$ number of N-atoms to an existing Cu$_n$ cluster, as $\Delta_c^0$ = -[E(Cu$_n$N$_n$)-E(Cu$_n$)-$n$E(N)].  Similarly, the energy gain $\Delta^0_m$ for the mono doped systems corresponds to the energy gain for a single N atom doping. First, the energetics of the pure Cu$_n$ clusters have been revised by analyzing their binding energies which are calculated as $E_b$ = -[E(Cu$_n$)-$n$E(Cu)]. Our calculated binding energies for the optimized pure Cu$_n$ clusters are $E_b$ = 2.24 eV, 3.66 eV, 6.33 eV, 8.57 eV and 11.44 eV for $n =$ 2, 3, 4, 5 and 6 respectively. The calculated values of $\Delta^0_c$ and $\Delta^0_m$ for the optimal structure of the respective capped and doped systems are shown in Fig. \ref{mes}. It is seen that $\Delta^0_c$ values increase monotonically 
with  increasing cluster size, which demonstrates increasing stability of the capped systems. On the 
other hand, the $\Delta^0_m$ values for the N-mono doped systems, show an overall decreasing trend 
with increasing cluster size. However, these values of $\Delta^0_m$ are always $+$ve for all the mono 
doped systems, indicating mono dopings are also energetically favorable. In order to understand the enhancement in stability, specially of the capped systems, we have analyzed the nature of the Cu$_2$ and CuN dimers bonding as well as the overall variation in the average Cu-Cu and Cu-N bond lengths in the optimized structures of the pure, mono-doped and capped systems as shown in Fig. \ref{bl}. In the atomic structure of a copper atom, the valence 3$d$ shell is completely filled with a single 4$s$ electron outside it. Therefore, two Cu atoms in a Cu$_2$ dimer, are coupled by week $\sigma$ bond resulting in a larger bond length of 2.22 \r{A}. Note that the estimated bond length of the Cu$_2$ dimer is in excellent agreement with its experimental value.\cite{dimer} When a nitrogen atom is attached to form a CuN dimer, the $s$-$d$ hybridized electrons of the copper atom interacts with the 2$p^3$ electrons of nitrogen atom resulting in a stronger bond and thereby a shorter bond length of 1.75 \r{A} for a CuN dimer. The charge density contour plot of the Cu$_2$ and CuN dimers as shown in the Fig. \ref{bl} demonstrates this type of bonding. Fig. \ref{bl} also shows that the average Cu-Cu bond lengths is larger in the mono-doped as well as capped systems compared to that in the pure Cu$_n$ clusters. It indicates that the Cu-Cu bond strength decreases upon the capping/doping. The enhanced stability of the capped systems, is therefore mainly caused by the enhanced stability of the Cu-N bonds, as the number of Cu-N bonds in the ground state structures of the capped systems, increases with increasing cluster size.

%============================================ Table 1 ======================================================================== 
\begin{table*}[!t] 
\caption{\label{tab:bemag} {\textcolor{red}{Total spin}, energy difference, $\triangle E$ with respect to the ground state, average values of Cu-N bond-length and Cu-Cu-N bond angle, distribution of magnetic moments for the ground state and first isomer of the Cu$_n$N$_n$ clusters. The average magnetic moments within parenthesis, correspond to $d$-orbital contribution for Cu-atom and $p$-orbital contribution for N-atom.}}
{\begin{tabular}{cccccccccccc} 
\hline
\hline
Cluster &   \textcolor{red}{Total}          &  & $\triangle E$ & & $\langle Cu-N\rangle$  & $\langle Cu-Cu-N\rangle$ & \multicolumn{3}{c} {Average spin ($\mu_B$/atom)} & \\
        &   \textcolor{red}{spin}   &  &   (eV)        &  & ($\AA$)                 &                          & \multicolumn{3}{c} {at the site of}       & \\
\cline{8-10}
        &    ($\mu_B$)     &  &               &   &                      &                          &          Cu  &  &    N                      &  \\
\hline
Cu$_3$N$_3$ &  8           &  &    0.00      &   &  1.78                &    150$^\circ$            & 0.43 (0.29)  &   &  1.39 (1.35)             &  \\
            &  4           &  &    0.20      &   &  1.77                &    150$^\circ$            & 0.31 (0.20)  &   &  0.66 (0.64)             &  \\
Cu$_4$N$_4$ &  10          &  &    0.00      &   &  1.78                &    145$^\circ$            & 0.35 (0.26)  &   &  1.36 (1.32)             &  \\
            &  10          &  &    0.24      &   &  1.77                &    135$^\circ$            & 0.36 (0.27)  &   &  1.37 (1.34)             &  \\
Cu$_5$N$_5$ &  12          &  &    0.00      &   &  1.77                &    140$^\circ$            & 0.32 (0.25)  &   &  1.36 (1.32)             &  \\
            &  12          &  &    0.25      &   &  1.77                &    137$^\circ$            & 0.31 (0.25)  &   &  1.35 (1.32)             &  \\
Cu$_6$N$_6$ &  14          &  &    0.00      &   &  1.78                &    135$^\circ$            & 0.29 (0.24)  &   &  1.31 (1.28)             &  \\
            &  14          &  &    0.15      &   &  1.78                &    136$^\circ$            & 0.29 (0.24)  &   &  1.31 (1.29)             &  \\
\hline
\hline

\end{tabular} }
\end{table*}
%=====================================================================================================================================

%############## Figure 3 ##############
\begin{figure*}
\rotatebox{0}{\includegraphics[height=7.5cm,keepaspectratio]{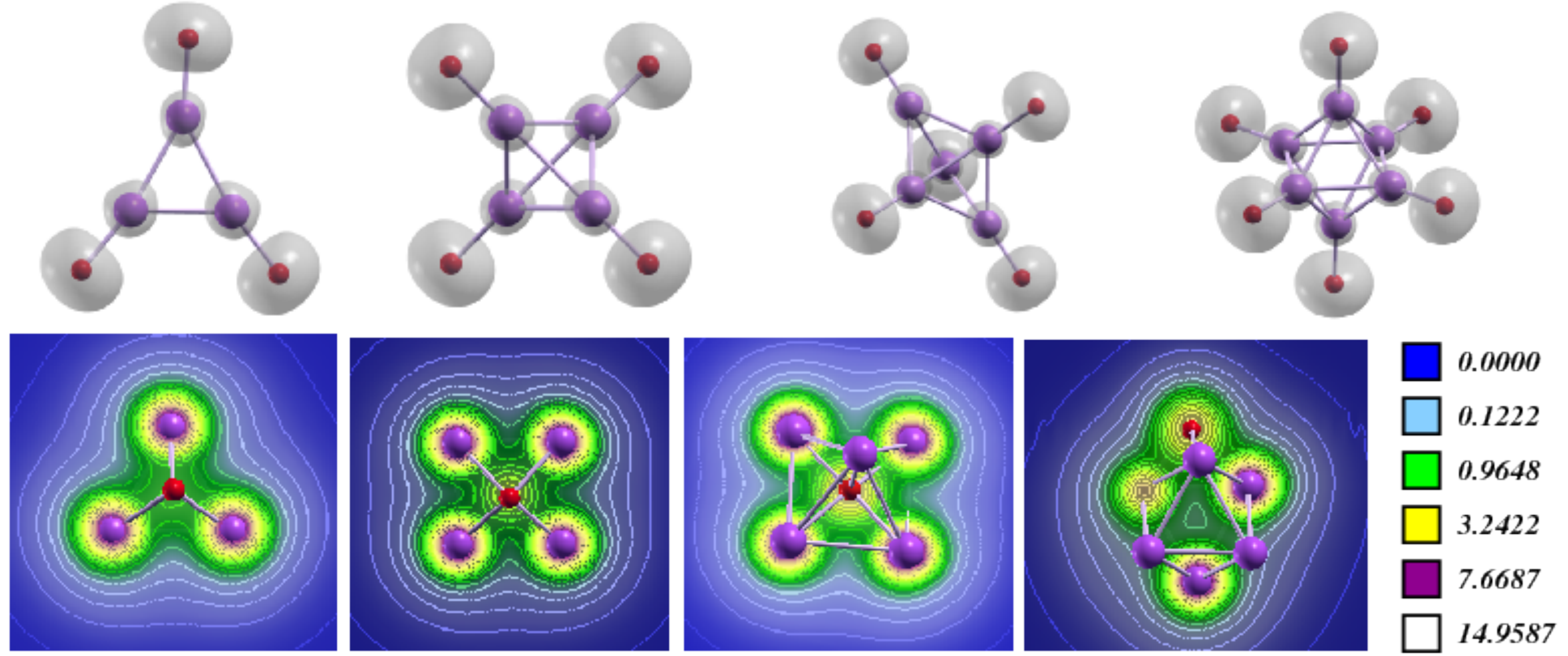}}
\caption{(Color online) Plot of spin density surface (upper panels) of the capped systems, Cu$_n$N$_n$ and charge density contours projected on to the xy plane (lower panels) for the mono doped systems, Cu$_n$N in their ground state with n = 3 $-$6. The scale for charge density which is shown aside in the lower panel, is chosen to be the same for all the mono doped structures, while an isovalue of 0.1 e$^-$/\AA$^3$ is used for plotting the spin density in the upper panel.}
\label{density}
\end{figure*}
%######################################

Finally,  we address the the main issue of the magnetic behavior for the three different classes of systems. First, for the pure clusters, the magnetic moments show an even-odd alternation in accordance with the previous reports.\cite{odd-even} For each of the even numbered Cu$_n$ clusters, all valence electrons are paired, giving a closed shell electronic structure which results in a zero net magnetic moment. Conversely, each odd numbered Cu$_n$ cluster, has one unpaired electron, which attributes a net magnetic moment of 1 $\mu_B$.  Our calculated magnetic moments for the ground state structures of the N-capped as well as N-monodoped systems, are given in Fig.\ref{mes}. It is seen that the magnetic moments of the mono-doped systems, also show an even-odd alteration, but with reversed period with respect to that of the corresponding pure systems. In other words, a N@Cu$_n$ cluster has net magnetic moment of 1 $\mu_B$ for the even values of $n$ and zero for the odd values of $n$. Considering the case of the ground state structure of the Cu$_2$N cluster, there is some charge transfer from N to Cu atoms due to hybridization between N-2$p$ and Cu-$sd$ orbitals. This results into net magnetic moment of 1 $\mu_B$ which is solely contributed by the N atom. Similarly for the MES of Cu$_3$N cluster, the $s$-electron of each Cu atom gets paired with a $p$-electron of the N-atom as there is significant overlap between the Cu and N atoms (cf. Fig. \ref{mes}), which now results into its zero net magnetic moment. The trend in the net magnetic moments in the subsequent Cu$_n$N clusters with $n >$ 3, would be readily understandable. This is because of the fact that the systems having odd number of Cu atoms  in excess of 3, such as for the Cu$_4$N and Cu$_6$N clusters, each also has one unpaired $s$ electron, which gives rise to 1 $\mu_B$ net magnetic moment. Conversely, all the valence electrons are paired for a mono-doped system having even number of Cu atoms in excess of 3 Cu atoms such as Cu$_5$N cluster and it results again zero net magnetic moment. Note this finding of the low magnetic moments of the N-monodoped Cu$_n$ clusters, is in direct contrast with the case of N-monodoped Mn$_n$ clusters, where giant magnetic moments have been predicted.\cite{n@mn}

In contrast to the mono-doped systems, the optimal N-capped Copper clusters show significantly enhanced magnetic moments. Among the capped systems, first the optimized Cu$_2$N$_2$ cluster has a special structure where each Cu atom is directly connected with both the N atoms and fulfills the perfect charge balance which results a zero net magnetic moment for it. However, the ground state structures of the other Cu$_n$N$_n$ clusters have net magnetic moments of 8 $\mu_B$, 10 $\mu_B$, 12 $\mu_B$ and 14 $\mu_B$ respectively for $n =$ 3$-$6. We find that the enhanced magnetic moments of the capped systems, result from the ferromagnetic coupling among the all N-atoms as well as Cu-atoms centered magnetic moments. In the N-capped Cu$_n$ clusters, Cu atoms become spin polarized because of charge transfer among the neighboring atoms. This spin polarization further magnetizes $p$-electrons of the N-atoms through $p$-$d$ hybridization. This hybridization in turn renders a ferromagnetic coupling state among all the constituent atoms. In order to understand the microscopic origin of the enhanced magnetic moments of the N-capped systems, we have performed the Mullikan population analysis of spins.\cite{mulli} In Table \ref{tab:bemag}, we provide information on the magnetic moments at the Cu and N sites for the ground state as well as first isomer of the N-capped clusters. We have also listed in Table \ref{tab:bemag}, their total spin, energy difference measured with respect to the ground state energy, values of $\langle Cu-N\rangle$ bond-lengths and $\langle Cu-Cu-N\rangle$ bond angles. The Mullikan analysis yields an averaged spin magnetic moments of 0.3$-$0.4 $\mu_B$ at the Cu site and 1.3$-$1.5 $\mu_B$ at the N-site. The important point to note is that the moments at Cu-sites arise mainly from Cu-$d$ orbital and from N-$p$ orbital in case of the N-sites. However, the major contribution to the enhanced magnetic moments of the N-capped clusters, arises from the N atoms. Also note that the averaged magnetic moments at Cu sites of the ground state structures of the capped clusters, though very small, show an overall decreasing trend with increasing cluster size. It, therefore, allows us to go one step forward in generalizing the net magnetic moments of the ground state structure of the four N-capped clusters by a formula 2$+$2$x$, where the first `2' denotes the magnetic moment contributed by the core of Cu atoms, while `2$x$' is the magnetic moment contributed by the $x$ number of N atoms attached as capping. This is a reasonable approximation because the N-capping allows charge transfer from N-atoms to the Cu-atoms and induces some magnetic moments to the Cu atoms totaling around 2 $\mu_B$ for each capped system. For better understanding of magnetic coupling, we have also plotted in Fig. \ref{density} the spin density surface corresponding to the ground state of the Cu$_n$N$_n$ clusters in the upper panel and charge density contour plot corresponding to the ground state of the Cu$_n$N clusters in the lower panels for the size range $n =$ 3$-$6. It again clearly indicates the ferro magnetic coupling of the constituent atom-centered moments for the capped systems and enhanced hybridization tendency for the mono doped systems which results in their tiny net magnetic moments.

 We note that this enhanced ferromagnetism of the N-capped Cu$_n$ clusters, has a closed relevance to the observed ferromagnetism of a Cu-doped GaN system. This is due to the fact that the $p$-$d$ hybridization mechanism is identified as the main factor for the ferromagnetic coupling of Cu-dopants in GaN.\cite{gan} We have carried out separate studies of the magnetic coupling between two Cu atoms in a GaN crystal using a 32 atoms supercell. Two Cu atoms were substituted at different Ga sites considering the various possibilities such as the near/far separation of the two dopants as well as their ferromagnetic/anti-ferromagnetic couplings. We found that in the MES of the 2Cu@GaN system, the coupling to be ferromagnetic with magnetic moments of 0.13 $\mu_B$ at each Cu site.

In order to verify our GGA-DFT results discussed above, we have also studied the small N-capped Cu$_n$ clusters using hybrid DFT functional. We used a screened form of hybrid functional HSE06 proposed by Heyd, Scuseria and Ernzerhof with screening parameter 0.206.\cite{hse} Note that the hybrid HSE06 functional has been used recently to correct GGA-DFT predictions of formation energy of transition metal oxide crystals\cite{hsecrys} as well as of magnetic interaction for transition metal clusters.\cite{hseclus} We find that the overall trend of enhanced magnetic moments of the N-capped Cu$_n$ clusters persists as found in the case of PBE calculations. In HSE calculations, we find that the N-atom centered magnetic moments get slightly increased and Cu-atom centered magnetic moments decrease slightly. This is attributed to the less hybridization between N-$p$ and Cu-$d$ orbitals. This is also reflected in slight increase of $\langle Cu-N\rangle$  bond-lengths of the capped clusters in the HSE calculations. It is thereby confirming that our result of the enhanced magnetic moments of the N-capped Cu$_n$ clusters is very robust irrespective of the choice of DFT functionals. It is therefore, also very desirable to look for its experimental verification. Recently, molecular beam deflection experiments have been employed to study magnetic moments of several capped transition metal clusters.\cite{knickel,knickel2} Such experiments would be useful in this respect.

\subsection{\label{endo}Capping and endohedral doping in Cu$_{13}$ cluster}

To study the effect of endohedral doping in a N-capped cage-like Cu cluster, we consider a particular case of a Cu$_{13}$ cluster. First, to determine the MES of a pure Cu$_{13}$ cluster, we have considered three most probable starting guessed structures, namely icosahedral (ICO), hexagonal bi-layered (HBL) and buckled bi-planar (BBP) structures. For the less symmetric HBL/BBP structures, all the 13 Cu atoms lie on the surface, thereby forming a cage-like structure. The morphology of an icosahedral Cu$_{13}$ cluster can also be regarded as a cage of the twelve surface Cu atoms, encapsulating another Cu atom at the center of the cage. It is to be noted that previous works on small copper clusters indicate a rich variety of structures and the icosahedral symmetry based structure is predicted not to be the true ground state structure for a Cu$_{13}$ cluster.\cite{planar,cun} Our calculated total binding energy and magnetic moment for the locally optimized three structures of the Cu$_{13}$ cluster, are 29.05 eV and 5 $\mu_B$ respectively for the optimal ICO structure, 29.53 eV and 1 $\mu_B$ respectively for the optimal HBL structure and finally, 29.65 eV and 1 $\mu_B$ respectively for the optimal BBP structure. So it is the optimal BBP structure of a pure Cu$_{13}$ cluster which appears as its MES in our calculations. Thereafter, each of the three structures was capped with N-atoms and we explored the potential energy surface to look for the MES of the N-capped Cu$_{13}$ cluster. To cap a HBL/BBP structure of the Cu$_{13}$ cluster, total 13 N atoms have been used to bind with all its 13 surface Cu atoms. On the other hand, total 12 N atoms have been attached with its 12 surface Cu atoms to cap the icosahedral Cu$_{13}$ cluster. Then each structure of the N-capped Cu$_{13}$ cluster, has been optimized for all possible spin multiplicities. After structural optimization, it is seen that the estimated values of $\Delta^0_c$/N-atoms are 1.81 eV, 1.64 eV and 1.65 eV for the optimal ICO-Cu$_{13}$N$_{12}$, BBP-Cu$_{13}$N$_{13}$ and  HBL-Cu$_{13}$N$_{13}$ clusters respectively. It is therefore, unambiguously showing that the N-capping to a Cu$_{13}$ cluster, favors an icosahedral Cu$_{13}$ core.

It is also interesting to note that the optimized structure of the icosahedral Cu$_{13}$N$_{12}$ cluster has a total magnetic moment of 29 $\mu_B$ which is very large compared to that of the pure optimal Cu$_{13}$ cluster and therefore, the sensitivity of enhanced magnetic moments of a N-capped Cu$_n$ cluster to cluster size, is beyond questionable. Similar to the smaller N-capped Cu$_n$ clusters, the large magnetic moment of the Cu$_{13}$N$_{12}$ cluster is contributed mainly by the spin polarized N atoms which carry in this case an average moment of around 1.4 $\mu_B$/N-atom. We note that there is a net charge transfer from the N atoms to the Cu atoms, which creates partial occupancy in the 3$d$-orbital of the Cu atoms and thereby, inducing spin polarization to them. Mulliken population analysis shows that the N-capping in the icosahedral Cu$_{13}$ cluster creates a local magnetic moment of about 0.25 $\mu_B$/atom to the surface Cu atoms. Above all, the spin-polarized N atoms are ferromagnetically coupled among themselves and also with the Cu-atom centered moments which results the giant net magnetic moment.

%############## Figure 4 ##############
\begin{figure}
\rotatebox{0}{\includegraphics[height=7.5cm,keepaspectratio]{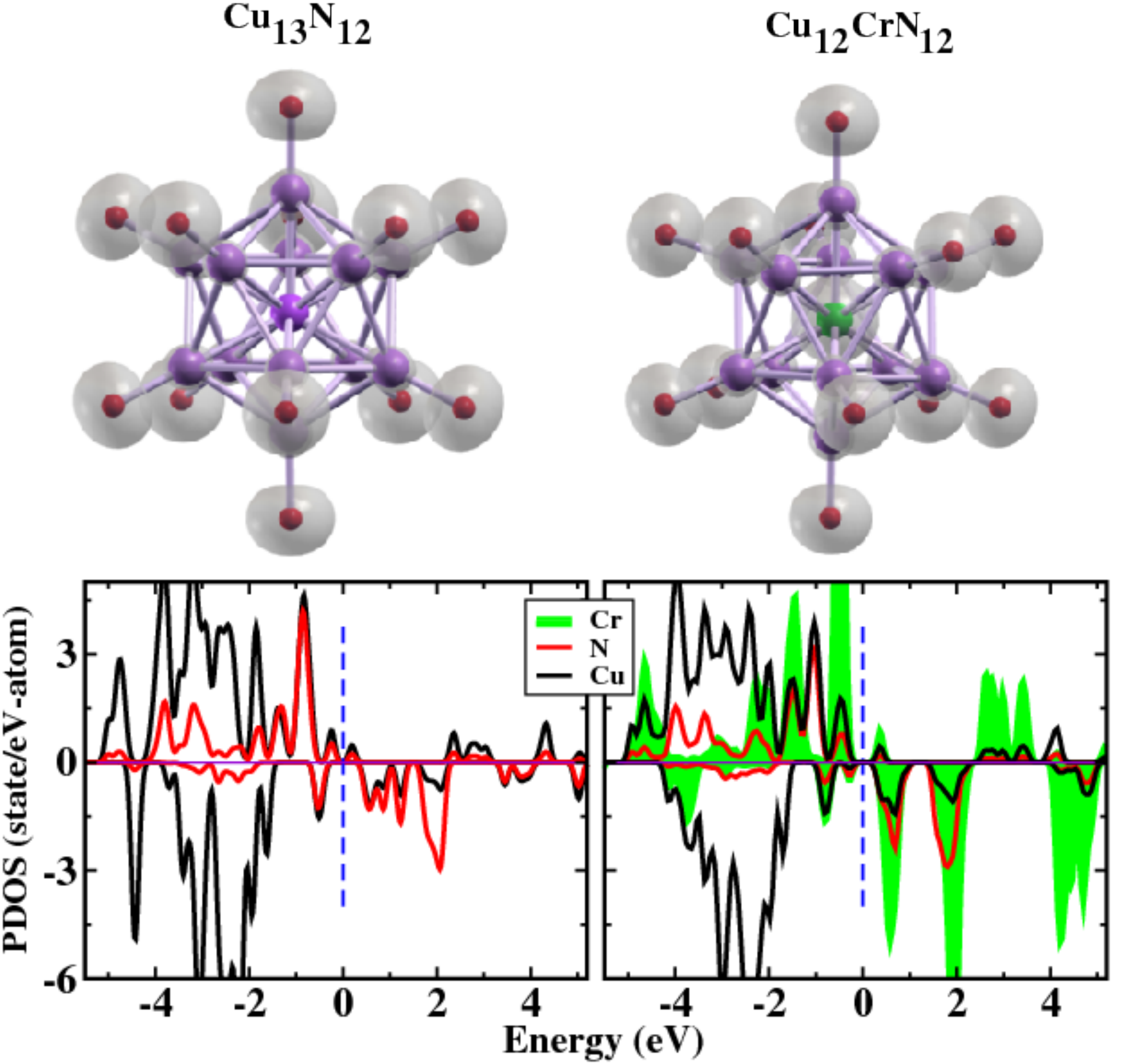}}
\caption{(Color online) Plot of spin density surface (upper panel) and projected density of states (lower panel) for the ground state structures of Cu$_{13}$N$_{12}$ and Cu$_{12}$CrN$_{12}$ clusters. The green colored ball in the Cu$_{12}$CrN$_{12}$ cluster structure, represents the Cr atom. Isovalue of 0.1 e$^-$/\AA$^3$ and smearing width of 0.1 eV are used in the plots of spin density and PDOS respectively. }
\label{pdos}
\end{figure}
%######################################

%############## Figure 5 ##############
\begin{figure}
\rotatebox{0}{\includegraphics[height=4.5cm,keepaspectratio]{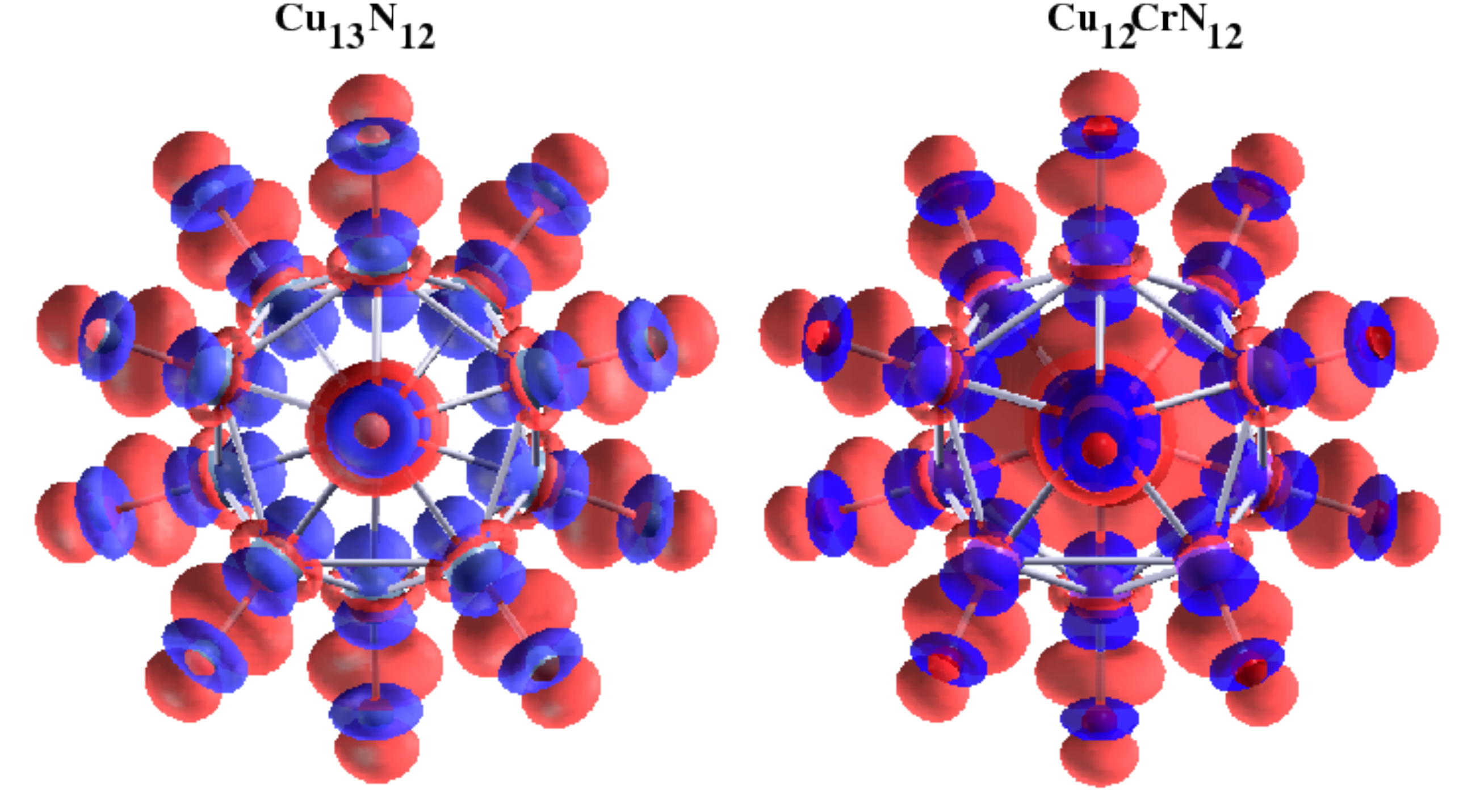}}
\caption{(Color online) Plot of charge transfers, $\Delta \rho$  for the ground state structures of Cu$_{13}$N$_{13}$ and Cu$_{12}$CrN$_{13}$ clusters. The isovalue is chosen at 0.03 e$^-$/\AA$^3$. The red (blue) colored surface represents the accumulated (depleted) charge. }
\label{chg_transfer}
\end{figure}
%######################################

The interesting point is that the magnetic moment of the Cu$_{13}$N$_{12}$ cluster can be further improved by the substitution 
of the central Cu atom by a Cr atom. Recently, the role of Cr atom as an endohedral dopant for constructing molecular units of 
highly magnetic functionalized materials, has been investigated for hydrogenated Si fullerenes.\cite{endo} With this in 
view, we have explored the effects of Cr substitution at one Cu-site in the icosahedral Cu$_{13}$N$_{12}$ cluster. We find that the 
optimized Cr-doped cluster prefers to have the Cr-substitution at the central site and it possesses a higher net magnetic 
moment of 34 $\mu_B$ along with a substantial energy gain of $\Delta_c^0$ = 18.1 eV. It is also important to note that a
 optimized bare Cu$_{12}$Cr cluster has zero net magnetic moment. Therefore, the N-capped Cr-encapsulated Cu$_{13}$ cluster
 shows significant improvement in magnetic moment as compared to that of the bare pure Cu$_{13}$ cluster as well as 
Cu$_{12}$Cr cluster. Though the magnetic moments of the Cu and N atoms are enhanced after Cr-substitution, it is the moment of 
the Cr impurity which contributes the most to the extra magnetic moment of the Cu$_{12}$CrN$_{12}$ cluster as compared to that of the 
Cu$_{13}$N$_{12}$ cluster. Moreover, all the atomic moments are ferromagnetically aligned. Fig. \ref{pdos} shows the spin 
density plots for the optimized structures of the Cu$_{13}$N$_{12}$ and Cu$_{12}$CrN$_{12}$ clusters. It clearly indicates
 the ferromagnetic alignments of the constituents in the two systems. Further, to understand the effects of N-cappings on the magnetic properties
of the pure Cu$_{13}$ cluster and the endohedral Cr-doped Cu$_{13}$ cluster, we have also plotted in the
 Fig. \ref{pdos}, the projected density of states (PDOS) of the different species of atoms in the two systems. Comparing the 
plot of the PDOS of the optimized Cu$_{13}$N$_{12}$ and Cu$_{12}$CrN$_{12}$ clusters, we see that there is hardly any 
 changes in the Cu-PDOS and N-PDOS between the two systems. For the Cr-PDOS of the Cu$_{12}$CrN$_{12}$ cluster, there is, however,
 significant difference between its majority and minority spin channels, which indicates the large magnetic moment character
 of the Cr atom. Mulliken population analysis of spins shows that the unpaired electrons of the central Cr-atom, indeed 
reside predominantly with it and do not appreciably  hybridize with the cage atoms. This is also reflected in the lower 
value of $\Delta_c^0$ for the optimal Cu$_{12}$CrN$_{12}$ cluster than that of the optimal Cu$_{13}$N$_{12}$ cluster. 
The interaction of the central Cr-atom with the cage-atoms seems largely mediated via the Cr-4$s$ state, which is possibly 
also manifested by the observed symmetric position of the Cr atom right at the center of the optimized cage. 

	To have a clear understanding of the nature of interactions of the central atom with the surrounding 12 Cu atoms and 
also that of the capping agents, N atoms with the surface Cu atoms, we have calculated the charge density difference, $\Delta\rho$ as
\[\Delta\rho = \rho({\rm C}_{12}{\rm CrN}_{12})-\rho({\rm Cu}_{12})-\rho({\rm Cr})-\rho({\rm N}_{12})\]
where $\rho($Cu$_{12}$CrN$_{12})$ is the total 
charge density of the optimized Cu$_{12}$CrN$_{12}$ cluster and the rest are charge densities of the isolated atoms for 
each species. The $+$ve value of $\Delta\rho$ indicates accumulated charge and $-$ve value for the depleted charge. 
Analysis of $\Delta\rho$, therefore, gives us clues about the overall charge redistribution after introducing the Cr atom 
at the central site. Fig. \ref{chg_transfer} shows the plot of $\Delta\rho$ for the MESs of the Cu$_{13}$N$_{12}$ and 
Cu$_{12}$CrN$_{12}$ clusters. It clearly indicates more accumulation of charge at the central Cr site of the 
Cu$_{12}$CrN$_{12}$ cluster, implying its high magnetic character and weak interaction with the surface Cu-atoms. 
Our Mulliken population analysis shows that it results in a Cr-atom-centered magnetic moment of around 3.2 $\mu_B$ and it is predominantly contributed by its $d$ orbital.

\section{\label{conclu}Summary and Conclusions}
In summary, N-capping is more promising than N-monodoping in  inducing giant magnetic moment in a Cu cluster. The N-capping is associated with enhanced hybridization of the valance $p$ orbitals of N atoms with the $s$-$d$ hybridized orbitals of the Cu atoms and it induces spin polarization to the Cu atoms. The ferromagnetic coupling among the constituent atoms gives rise to the giant magnetic moments for the capped systems. This study also shows that the magnetism of the N-capped cage-like structures of a Cu cluster, can be enhanced further upon Cr encapsulation. As the optimal Cu$_{12}$CrN$_{12}$ cluster possesses both enhanced hybridization among the valence orbitals of the constituent atoms as well as larger net magnetic moments, a Cr-encapsulated cage-like structure of N-capped Cu cluster, may be a better candidate for dilute magnetic semiconductors and experimental investigations in this direction would be exciting.
\acknowledgments
 We are grateful to Prof. T. Saha-Dasgupta for providing the computational facilities as well as for many simulating discussions. S.D. thanks Department of Science and Technology, India for support through INSPIRE Faculty Fellowship, Grant No. IFA12-PH-27.
%\section*{References}

\end{document}